\documentclass[12pt]{article}
\usepackage{times}

\usepackage[toc,page]{appendix}
\usepackage{booktabs}
\usepackage{enumerate}
\usepackage[shortlabels]{enumitem}
\usepackage{fullpage, epsfig, subfigure, wrapfig}
\usepackage{wrapfig, amsmath,amssymb, algorithm, algorithmic}
\usepackage{multirow, tabularx, indentfirst, color, cite}
\usepackage{hyperref}
\hypersetup{
  colorlinks   = true, 
  urlcolor     = black, 
  linkcolor    = black, 
  citecolor   = black 
}
\usepackage{cleveref}

\usepackage[compact]{titlesec}
\usepackage{palatino}

\titleformat{\section}
  {\normalfont\fontsize{14}{19}\bfseries}{\thesection}{1em}{}
\titleformat{\subsection}
  {\normalfont\fontsize{12}{17}\bfseries}{\thesubsection}{1em}{}
\titleformat{\subsubsection}
  {\normalfont\fontsize{12}{17}\selectfont}{\thesubsubsection}{1em}{}

\usepackage{geometry}
\makeatletter
\newcommand\figcaption{\def\@captype{figure}\caption}
\newcommand\tabcaption{\def\@captype{table}\caption}
\makeatother

\begin{document}
\begin{titlepage}
\begin{center}
\makebox{}
\vspace{5cm}



{\LARGE \bfseries From Texts to Shields: Convergence of Large Language Models and Cybersecurity} 

\vspace{2cm}

\begin{center}
    Tao Li, Ya-Ting Yang, Yunian Pan, and Quanyan Zhu \\
\end{center}

New York University
\vfill

\centering \today 
\end{center}
\end{titlepage}

\newpage
\tableofcontents

\newpage

\title{\bf \Large From Texts to Shields: Convergence of Large Language Models and Cybersecurity}
\author{Tao Li, Ya-Ting Yang, Yunian Pan, and Quanyan Zhu\\ 
Department of Electrical and Computer Engineering\\ New York University\\
\texttt{\{tl2636,yy4348,yp1170,qz494\}@nyu.edu}
}
\date{}
\maketitle
\begin{abstract}
    This report explores the convergence of large language models (LLMs) and cybersecurity, synthesizing interdisciplinary insights from network security, artificial intelligence, formal methods, and human-centered design. It examines emerging applications of LLMs in software and network security, 5G vulnerability analysis, and generative security engineering. The report highlights the role of agentic LLMs in automating complex tasks, improving operational efficiency, and enabling reasoning-driven security analytics. Socio-technical challenges associated with the deployment of LLMs---including trust, transparency, and ethical considerations---can be addressed through strategies such as human-in-the-loop systems, role-specific training, and proactive robustness testing. The report further outlines critical research challenges in ensuring interpretability, safety, and fairness in LLM-based systems, particularly in high-stakes domains. By integrating technical advances with organizational and societal considerations, this report presents a forward-looking research agenda for the secure and effective adoption of LLMs in cybersecurity.
\end{abstract}

\section{Background and Scope}
The recent success in large language models (LLMs) has shown that they have the capacity to process and understand vast amounts of textual data, making them indispensable for natural language processing tasks \cite{wei2022emergent,min24llm-survey, tao24spo}. They play a major role in processing text, time-series data, and prediction \cite{bubeck23gpt}. These successes are particularly relevant to cybersecurity because they empower us to harness the power of human-like language understanding and predictive capabilities to enhance network security. In cybersecurity, LLMs can effectively analyze and interpret security reports, threat intelligence feeds, and logs \cite{ranade21ner,guo21detect, setianto22gpt-2c, huang23detect}. They assist in identifying emerging threats, vulnerabilities, and attack patterns, thereby strengthening threat detection and incident response \cite{jo22vulcan, alam23ladder, wu22its}. Furthermore, their ability to comprehend human-generated content in emails, chat logs, and social media posts enables the detection of phishing attacks, insider threats, and other security-related issues \cite{ranade21fake-CTI, maneriker21phishing-url, ayoobi23phishing-detection}.

Moreover, these models can be employed in developing advanced anomaly detection systems that continuously monitor network traffic, system logs, and user behavior \cite{wang2023tbdetector,ferrag2023revolutionizing}. By recognizing unusual patterns or deviations from the norm, they aid in early threat detection, providing security teams with valuable insights to respond proactively \cite{perrina2023agir, fayyazi2023uses, tao-qz24symbiotic}. Additionally, LLMs can generate and review security policies, patch management, and vulnerability scanning \cite{le2023log,das22vwx-bert}. By automating these processes, organizations can respond more efficiently to security incidents and reduce the workload on security professionals. This integration of large language models into the cybersecurity landscape represents a significant advancement in fortifying our digital defenses and staying ahead of evolving threats.

As the threats are growingly sophisticated and network systems are increasingly connected, there is a need to create disruptive technologies to transform the way we secure our networks. The convergence between network security and AI is critical for the nation’s cybersecurity and promises a secure and resilient digital future. This report discusses opportunities to reshape the landscape of network security through the integration of large language models (LLMs). Drawing on insights from the fields of network security, artificial intelligence, language modeling, and formal methods, it outlines a dynamic research agenda for advancing next-generation security technologies and practices.

This report synthesizes current research issues, key challenges, and future directions regarding the application of LLMs in network security. It draws on interdisciplinary perspectives from academia, industry, and government, and the discussions are organized into three primary focus areas:
\begin{enumerate}[(A)]
    \item LLMs for Software and Network Security;
    \item LLMs for Secure and Resilient Network Service and Management;
    \item Human and Social Aspects of LLMs in Cybersecurity.
\end{enumerate}

Each area has its fundamental challenges that need to be addressed. Area A focuses on the role of LLMs in malware detection, software testing, log analysis, penetration testing, and blue/red teaming, which can span a wide range of topics. The focus area B shifts the attention to network operation applications, which aim to address the questions of network operations, response, resiliency, forensics, and high-level service management. The focus area C tackles the human and social aspects of LLM for cybersecurity, aiming to address the concerns related to trust, ethics, human interactions, datasets, and the applications in misinformation and manipulation. Participants presented their research and engaged in in-depth discussions about the challenges and opportunities in their respective fields. 

As a summary of fruitful discussions at the workshop, this report lays out the transdisciplinary foundations and challenges in LLMs and their connections with security applications. This report provides an overview of the key topics and ideas discussed during the workshop, as well as the major research issues and challenges identified. By integrating perspectives from academia, industry, and government, the report offers actionable guidance for researchers, policymakers, and practitioners engaged in the design and implementation of next-generation cybersecurity solutions.

\section{LLM Applications in Network Security}
\label{sec:llm-network}



\subsection{LLM for Analytics of Cyber Threat Intelligence}

Many cybersecurity reports are long, unstructured, and not actionable, which creates difficulties for organizations in quickly responding to threats. It is observed that reports like cyber threat intelligence (CTI) often lack immediate usability, even though they are crucial for mitigating cyber threats. Prior efforts have worked on NLP applications to process these reports and make the information actionable. However, LLMs have significantly improved this process, enabling better extraction of relevant data and threat patterns from extensive reports \cite{ehab22bert}.

LLMs can play an important role in extracting actionable attack behaviors (e.g., adversary tactics, techniques, and procedures) from the vast amount of unstructured text \cite{ehab23semantic}. Yet, one must be aware of the challenges of semantic analysis in cybersecurity, where terminologies and context often differ from common language uses (e.g., ``kill'' meaning to stop a process). To address such challenges, LLM fine-tuning is necessary. For example, fine-tuned models, such as ``SecureBERT,'' are specifically designed to understand and process cybersecurity-related content \cite{ehab23securebert}. This has demonstrated improved accuracy in predicting vulnerabilities, mapping attack behaviors, and identifying potential threats.

While LLMs are highly useful in reducing the workload by automating the extraction and analysis of cybersecurity reports, they still have limitations. These include challenges in logical reasoning, prompt engineering, and fine-tuning to achieve specific tasks. It should be noted that experts are still required to design the prompts properly for effective outcomes. Another significant issue in cybersecurity is the lack of diverse and comprehensive attack data, making it hard to train models that can cover all threat scenarios \cite{tao23ztd}. Additionally, LLMs sometimes fail to provide accurate or reliable results, particularly when reasoning or understanding deeper context. 

Finally, cybersecurity goes beyond data analysis—it is about making decisions that require clear understanding, context, and reasoning, areas where LLMs, though improving, still need development.

\subsection{ LLM for 5G Network Security}

5G technology drives innovation across various fields. However, it also presents significant security challenges, including vulnerabilities due to the complex nature of the protocols and the lack of encryption in certain communications \cite{zhu13gamesec}. While mobile technology continues to evolve \cite{yin-tao24pirl, tao25dt-pirl}, significant security gaps remain. One of the primary issues discussed was the vulnerability in communication systems before encryption is established. Attackers can exploit these vulnerabilities to compromise private communications or launch malicious attacks \cite{li2024decision}. To ensure 5G security, one must address protocol-level vulnerabilities and enhance system-level security.

LLMs display a great potential for both protocol-level testing \cite{meng2024large} and system-level security design \cite{deng24pentest}. As for testing, there are two major approaches that LLMs can be part of. One is the top-down approach, which involves understanding the design and specification of the 5G system (which is often described in voluminous documentation over 1,000 pages of protocol descriptions). Test cases are generated from the formal specifications and protocols. While effective, this approach is resource-intensive, requiring a significant amount of manual effort to parse these massive documents and identify the necessary components for security testing. The other one is bottom-up. Unlike the top-down approach, this method focuses on analyzing network traffic data. Instead of manually interpreting complex protocol descriptions, this method allows testers to directly examine the network traffic to detect anomalies or security flaws. It should be noted that this approach is more adaptable because even if the system changes, testers can continue analyzing the traffic without needing to understand the underlying protocol fully.

LLMs can play an instrumental role in 5G security, particularly in the following aspects:
\begin{itemize}
    \item automating the generation of test cases;
    \item extracting protocol rules and constraints from large datasets;
    \item creating new models for fuzzing (a testing method) and penetration testing;
    \item offering reasoning capabilities to refine and optimize security testing.
\end{itemize}
Recent research has presented examples of how LLMs are being used to generate grammars from protocol descriptions and then create test cases based on those grammars. This automation helps in testing large systems and identifying vulnerabilities that may have been missed by manual testing processes \cite{hongxin2024automated}.  LLMs can also help in traffic fuzzing by generating mutated traffic patterns to test the robustness of 5G systems. This allows testers to evaluate how the system responds to abnormal or unexpected inputs, helping uncover security flaws in different layers of the 5G architecture. Another example was their penetration testing tools built using machine learning models. These tools can conduct penetration tests based on generated test cases, providing a more thorough examination of security vulnerabilities in 5G systems.

In conclusion,  one must acknowledge the importance of using advanced technologies like LLMs to address the growing complexity and security risks in 5G networks. LLMs provide a way to automate many aspects of security testing, reducing the reliance on manual effort and allowing for faster, more effective detection of vulnerabilities. LLMs, in combination with traditional security practices, could revolutionize how we approach security testing in the future, especially for large, complex systems like 5G.

\subsection{Generative Security Application Engineering}

Generative AI and LLMs are revolutionizing cybersecurity practices. In addition to research efforts on LLM-based cybersecurity technologies, another question of equal importance is how future engineers can be supported in developing reliable, safe, and cost-effective generative security applications.

Toward this objective, a pilot course has been designed and taught, which covers code summarization, reverse engineering, vulnerability discovery, threat intelligence, social engineering, and code generation. In this course, students are instructed to evaluate and benchmark various models, experiment with prompts, and build secure applications using LangChain and security frameworks. The following are six lessons learned from this course. 
\begin{itemize}
    \item Creativity Matters: The application of LLMs in cybersecurity requires creativity and imagination beyond basic mechanics.
    \item Portability Matters: LLM models should be chosen based on the task (e.g., reasoning, code generation) while balancing between different models and contexts.
    \item Skepticism Matters: Beware of cognitive biases (e.g., survivorship bias). LLM models can provide plausible but incorrect outputs, emphasizing the need for verification.
    \item Change Matters: Technologies and LLM models evolve rapidly, and engineers need to adapt with an agile mindset.
    \item Safety Matters: It is necessary to address vulnerabilities in LLMs, such as prompt injection attacks and insecure plugin configurations.
    \item Costs Matter: time, energy, and financial costs in LLM applications are significant. Proper human interventions can mitigate mistakes made by LLMs and help reduce these expenses.
\end{itemize}

In conclusion, the final lesson is that teaching matters. Students should be encouraged to explore and innovate LLM usage, applying generative AI to automate a range of security tasks while being cautious of the fast-evolving landscape and security challenges.

\subsection{Discussion}
\begin{enumerate}[1)]
    \item {\it  How can LLMs be made trustworthy in the cybersecurity context?}
    
    LLMs are infamous for their inconsistency issues. When taking the same input, LLMs can generate totally different outputs, which causes trust issues and prevents the actual deployment of LLMs in mission-critical scenarios. Unfortunately, this inconsistency issue is due to its complicated inner workings that are hidden from the public, and hence, solving the issue from its root seems impractical. One viable solution is to resort to an additional verification process by installing certain programs or even another LLM. 
 
    \item {\it  Multi-modal multi-task general-purpose LLMs v.s. Domain-specific LLMs: Which is more practical when applying LLMs to cybersecurity tasks?}

    The latter looks more promising. First, one has to be aware that cyber defensive operations involve multiple tasks of different kinds. Each of them presents data with various structures that are beyond any LLM model. Hence, it is not promising to train a generalist model to handle all these data. Second, training a general-purpose model requires an astronomical amount of data, which is unlikely to be accessible in cybersecurity contexts. As organizations enjoy distinct IT infrastructures and face different adversarial threats, their security data are highly unstructured and heterogeneous, which may not be suitable for LLM training. Moreover, such data are often private, and organizations may not be willing to share their data to create a data pool with enough amount of datapoints that support LLM training. 
 
\end{enumerate}

\section{LLM Agent and Applications}
\label{sec:llm-agent}



\subsection{ The Rise of the Planet of the Agent}

AI-powered agents are increasingly being used in a variety of entrepreneurial scenarios to transform business operations. These agents are becoming pivotal in streamlining processes such as incident analysis, where AI can quickly assess and summarize critical events, providing real-time insights to decision-makers. Similarly, conversational assistants powered by AI are enhancing human interactions, offering context-aware suggestions, and facilitating smoother communication. Moreover, AI agents excel in data extraction and knowledge building, where they automate the collection and structuring of information, leading to more efficient knowledge management systems. Another key area is search and contextual retrieval, where AI enhances traditional methods by quickly identifying and delivering the most relevant data, enabling businesses to navigate large datasets with ease and precision.

Knowledge-enhanced large language models (LLMs) are being proposed to handle increasingly complex and novel situations. These models are designed to perform well even in scenarios where they encounter information or questions they haven’t seen before, thanks to their ability to integrate external knowledge sources. This capability is particularly valuable in fields like cybersecurity, where threats and attack methods evolve rapidly, requiring AI systems to be adaptable and informed beyond just historical data.

A more advanced technique, Retrieval-Augmented Generation (RAG) \cite{lewis2020retrieval}, is also emerging as a solution for handling multi-hop queries—those that require the AI to connect various pieces of information across multiple steps to arrive at a comprehensive answer. In areas such as threat intelligence, where understanding the profile of a threat actor involves analyzing multiple data points, RAG breaks down information into smaller, manageable chunks. It then searches for similar patterns across these chunks to ensure accurate retrieval. However, basic RAG methods may not always be enough, which has led to the development of a Graph-RAG approach \cite{procko2024graph}. This method enhances the RAG process by incorporating knowledge graphs, where text chunks are converted into graph-based elements that can be linked and analyzed in communities. These communities provide a more structured representation of information, allowing the AI to generate summaries that reflect not only individual data points but also the relationships between them, offering richer and more actionable insights.

Emerging technologies are also enabling a new Agentic LLM-based approach, where LLMs are used as advanced reasoning engines. Unlike traditional systems where actions are hardcoded, these LLM agents are dynamic, allowing them to plan actions based on natural language input and interact with the external world in a manner similar to how humans operate. This capability is particularly useful in environments that require adaptability, such as in complex decision-making tasks where the AI must gather data, plan, and execute actions without being restricted to predefined behaviors. For example, in the context of using an Agentic LLM-based agent with function-calling capabilities, the user can first ask a question and specify the desired function signatures. The LLM then returns the appropriate tool names (functions) along with the necessary parameters for calling those functions. Once the user executes the function locally with the provided parameters and obtains the output, the LLM acknowledges the action, summarizes the output, and delivers a user-friendly response to the original question. This process allows for seamless interaction between the user and the LLM, with the LLM assisting both in finding the executing functions and interpreting the results. This combination of reasoning and interaction with human users makes AI agents significantly more versatile and capable of performing a wider range of tasks.

\subsection{ Cybersecurity AI Evolves Rapidly with LLMs and Agents}

The field of cybersecurity is evolving rapidly with the integration of large language models (LLMs) and intelligent agents \cite{tseng2024using}. A key question is: Why use LLMs in cybersecurity? Deep learning (DL) has traditionally been effective in addressing known cyber attacks, as it allows systems to scale by fine-tuning new data points and adapting to different programming languages. However, DL presents significant challenges, such as the high costs associated with cleaning and labeling data, issues with data imbalance, and difficulties in explaining results. Moreover, DL systems are prone to adversarial attacks \cite{li2024meta}. In the case of unknown threats, such as zero-day attacks, reinforcement learning (RL) can offer more dynamic defenses by making moving target strategies adaptive. However, RL systems have their own limitations, particularly when attacks are short-lived, leading to gaps in the knowledge they acquire.

In comparison, LLMs offer unique advantages for cybersecurity \cite{wang2023effectiveness}. Unlike DL, which has limited generalization across domains, LLMs are inherently cross-domain, enabling them to apply knowledge across a variety of fields. LLMs also benefit from self-supervised learning, allowing them to learn from vast amounts of data without needing domain-specific labeling, which is often required for DL models. Another critical feature of LLMs is their potential to develop reasoning capabilities, which are essential for in-depth security analysis. While DL models may achieve high accuracy, they often lack the capacity for reasoning and explanation. LLMs, especially when using methods like ChatGPT’s chain-of-thought reasoning, can break down complex, multi-step security problems into intermediate steps, providing more structured and logical responses.

In the debate between human-LLM collaboration versus autonomous agents in cybersecurity, a key issue is reliability. Many cybersecurity decisions are mission-critical, and it is unacceptable for AI agents to be misled by factual inaccuracies present in LLM outputs. For non-critical tasks, autonomous agents may be more suitable, though a human supervisor is still recommended. When designing such agents, the primary focus should not only be avoiding factual errors but also actively tackling and correcting them to ensure reliable decision-making processes. This balance between human oversight and agent autonomy is crucial for deploying AI in cybersecurity contexts.

For example, User Privilege Operation (UPR) remains a specific challenge in server-side programs. The analysis of UPR variables is crucial for identifying vulnerabilities, but heuristic methods like regular expressions are fundamentally limited in their effectiveness. A more promising approach is to leverage the power of LLMs, which can perform well in analyzing small-sized C functions. However, LLMs may still struggle with larger ones due to memory limitations. Hence, a proposed workflow \cite{wang2024hybrid} for agents analyzing UPR can involve several steps: constructing a program dependency graph (PDG), slicing the problem into manageable parts, examining specific code statements, and having the LLM evaluate those short statements. The final UPR score is derived from aggregating the ratings given to individual code statements, providing a more comprehensive understanding of potential vulnerabilities.

However, there are still higher-level challenges \cite{zhang2023featureslearnedcodebertempirical}. One of the most significant is that existing LLMs suffer from factual errors, which pose a serious problem when relied upon for AI agent analysis in cybersecurity. These errors can undermine the effectiveness of automated analysis, and while human intervention may help correct such issues, it introduces the risk of turning the process into a repetitive task, which reduces efficiency. Addressing factual errors without becoming dependent on repetitive human correction remains one of the most pressing issues in the development of reliable AI agents for cybersecurity.

\subsection{ Agent of Agents: Meta LLM-Agent for Security Operations}

Today’s security operations are largely manual, slow, costly, and often ineffective. Red Teams, which are tasked with simulating advanced persistent threat (APT) attacks \cite{singhal2023advanced}, rely heavily on skilled personnel to perform infrequent and in-demand assessments that often lack formal quantitative metrics. Meanwhile, Blue Teams are responsible for monitoring, scanning, and incident response, but these processes are often incomplete and expensive. As a result, there is growing interest in moving toward more autonomous agents that can handle security operations with greater efficiency and lower operational costs. These agents can make decisions autonomously, adapt to new threats, and provide real-time automated responses. This shift is part of a broader transition from traditional rule-based engines to more sophisticated AI-driven engines, such as those powered by Reinforcement Learning (RL) and Large Language Models (LLMs).

In the context of RL agents, the Purple Teaming approach stands out as a viable solution for simulating security environments. Within the Purple Teaming framework, a Red Team agent simulates APT attacks while a Blue Team agent learns and develops defensive strategies \cite{li24col-aisg, hammar-tao25col}. The Purple Agent acts as a mediator, transferring insights learned from these interactions to improve overall network defenses. This process allows for a more systematic and automated approach to developing defense strategies for the Blue Team. A case study on penetration testing illustrates how RL-based agents can streamline what is traditionally a manual and complex process, allowing for more efficient security operations. However, RL-based agents come with limitations. RL-based systems often require stylized modeling and additional data processing, and they struggle with certain long-standing challenges within the RL framework \cite{pan2023first,li2024meta}. This is where LLM agents come in, offering advantages that RL-based systems often lack. LLMs leverage pre-trained knowledge bases, enabling them to understand and generalize textual data across various domains. This generalization capability is particularly valuable in dynamic security environments where new threats continually emerge as attackers evolve \cite{tao23sce}. With these advantages, LLMs can offer a more flexible and adaptive solution for cybersecurity operations.

Take penetration testing as an example; past efforts have explored both RL-based and LLM-based approaches, but these systems are not fully autonomous and often require human intervention. This highlights the need for a more comprehensive agentic solution, one that moves beyond single-task agents to a system of multiple specialized agents, since no single agent can handle the full complexity of modern cybersecurity operations. This leads us to a multi-agent systematic solution. Each agent within the system can be specialized to handle specific tasks, leveraging its unique capabilities and resources. This principle applies not only to RL-based agents but also to LLMs and other AI models, where specialization and collaboration between agents can enhance overall performance \cite{tao24picol}.

Another viable approach that shares similar wisdom is the concept of the Meta Agent or the Agent of Agents. A Meta Agent is essentially a mosaic of specialized agents, each focused on particular tasks \cite{shutian23erm, tao22confluence}, resulting in a cost-efficient and customizable agentic solution. In automated penetration testing, the Meta Agent can be viewed as solving an optimization problem, such as maximizing its utility value while adhering to budget constraints. By synthesizing insights from various agents, the Meta Agent can tackle complex tasks that no single agent can manage on its own. In addition to the Meta Agent concept, the integration of game-theoretic methods \cite{pan2021efficient,panmasage, tao_info} and LLMs also provides a symbiotic framework for cybersecurity operations \cite{tao-qz24symbiotic}. From the bottom up, game models provide a strategic framework for analyzing and defining high-level goals. From the top down, LLMs can take these strategic commands and translate them into operational tactics, allowing the system to execute the desired actions effectively. This integration of strategy and operation offers a more holistic approach to managing security operations, ensuring that high-level decisions are carried out with precision at the tactical level.

There are still significant challenges to address in the field of security operations. One of the primary concerns is ensuring accountability and maintaining safety in autonomous systems. Additionally, achieving lifelong learning—where agents can continuously consolidate and transfer knowledge—is critical for maintaining their long-term effectiveness. Another challenge is ensuring online adaptation in nonstationary environments, where threats and attackers continuously evolve in real-time \cite{tao23cola}. Overcoming these issues is crucial for the successful development and deployment of AI-driven solutions in cybersecurity, ensuring they can respond dynamically to ever-changing security landscapes.

\subsection{ Smaller Language Models, More Robust}

In recent years, deep learning (DL) has become pervasive across numerous applications, yet it still struggles with several challenges. As we advance, the focus is shifting toward trustworthy language models (LMs) that are robust, smaller in size, and offer better interpretability. These characteristics are essential to ensure that models perform reliably in diverse and unpredictable real-world environments.
When discussing robustness, we refer to a system’s ability to reduce generalization error, which is how accurately a model predicts outcomes on unseen data. This concept is also known as out-of-domain robustness. The goal is to maintain consistent performance across different datasets while providing statistical guarantees about the model's reliability. Robustness can be defined as a system's capacity to certify, with high probability, that its performance will not deviate significantly when tested on data from different domains or exposed to various types of noise. One method for quantifying robustness is through a probability framework defined as $(\epsilon, \gamma)$-robustness \cite{yu-etal-2022-measuring}: $\text{Pr}(X-\mu<\epsilon) > 1-(\sigma^2/\epsilon^2)\gamma$. In this model, the robustness of a system is measured using parameters such as the human or automatic evaluation score ($X$), the average score across all test sets ($\mu$), and variance ($\sigma^2$) in the evaluation scores. A lower value of $\gamma$, the inverse robustness indicator, suggests a more robust model, meaning its performance is more consistent across different conditions. The goal is to ensure that the model can generalize well beyond the data it was trained on.

To improve both robustness and efficiency, one viable approach involves data filtering using Reinforcement Learning (RL) \cite{yu-etal-2023-probabilistic}. Here, the selection of data is framed as a Markov Decision Process, with the RL agent selecting batches of data to maximize the reward consisting of robustness, diversity, and uncertainty. This approach allows the system to become more selective and focused in its training, avoiding unnecessary data that might reduce performance. Key challenges include formalizing data selection as an optimization problem and designing reward functions that accurately reflect the desired outcomes, such as improving robustness. A critical element of this approach is the design of reward functions that guide the RL agent in selecting data. Important factors include dispersion, which captures the variability of the data; data uncertainty, which accounts for the unknown or unpredictable elements within the data; and robustness, which ensures the model maintains its reliability across different conditions. These factors are used to create a balanced reward system that encourages the model to filter high-quality, diverse data that strengthens its generalization ability.

In summary, the robustness of language models involves ensuring stable performance across varied test sets and conditions. Methods such as leave-one-out simulation can help measure this consistency. Improving robustness does not always require larger datasets—sometimes, selecting smaller, high-quality datasets leads to better performance. By employing RL-driven data filtering strategies, models can achieve greater efficiency and reliability, making them more adaptable to real-world applications while remaining lightweight and easy to interpret.

\subsection{ Discussion}
\begin{enumerate}[1)]
    \item {\it  What is the cost of building an agent using GPT?}
    
    Most practitioners are not solely interested in minimizing the cost per query or reducing the total number of queries. Instead, they are prioritizing optimizing the overall strategy to achieve the best outcome within a given budget. The emphasis is on achieving optimal results within resource constraints rather than calculating or strictly minimizing query costs using GPT.

    Current industry practices leverage substantial computational resources, with cost considerations posing minimal constraints. However, there is increasing interest in scaling down to smaller servers for microservices to improve resource efficiency in future system architectures.

    \item {\it Should we use smaller, specialized agents organized in a hierarchy or one highly powerful agent?}

    The optimal agent configuration depends on the specific use case. One approach involves deploying multiple smaller agents, each restricted to a specific set of tools or APIs; alternatively, a single, more capable agent with access to a broader range of resources may be utilized. A dynamic strategy can also be adopted, enabling agents to access different tools and concepts depending on the operational context. In some scenarios, it may be more effective to dynamically generate an agent tailored to a particular task or problem. Flexibility remains essential, with the choice between specialized and unified agents determined by the nature of the task and the specific user requirements.

    \item {\it What are the main challenges in building and deploying these agents?}

   A major challenge identified is ensuring consistency and establishing coherent benchmarks across diverse use cases, which is essential for maintaining reliability and accurately measuring performance. Another key challenge lies in the clear definition of tasks and the development of appropriate evaluation platforms and metrics; without well-defined tasks, optimizing agent performance and comparing outcomes becomes difficult.

    An illustrative example highlights the importance of task complexity. Efforts have been made to prioritize use cases that are relatively straightforward to analyze. For instance, in attack detection scenarios, analyzing initial access data, such as signing logs or email data, tends to be less complex due to the limited and well-defined data sources involved. In contrast, analyzing lateral movement within a network involves integrating multiple, heterogeneous data sources and maintaining memory context across multiple iterations, significantly increasing task complexity. Current efforts aim to systematically define and quantify task complexity by examining factors such as data volume, data diversity, and the need for deterministic workflows.

\end{enumerate}

\section{Socio-Technical Aspects of LLM and Security}
\label{sec:llm-human}



\subsection{ Human-Centric Security Training with LLMs}

Cybersecurity training is undergoing a crucial transformation: instead of prescribing static rules like spotting suspicious links or vetting email senders, the focus is shifting toward resilience-building experiences powered by large language models. By generating personalized phishing simulations that mirror individual behaviors, psychological triggers, language styles, and attack strategies, these systems can expose weaknesses in response patterns and gradually escalate the complexity of threats \cite{heiding2024evaluatinglargelanguagemodels,10466545}. Yet detecting malicious content is just the starting point---truly effective programs teach contextual awareness, social-cue recognition, and anomaly detection in tone and timing, equipping learners to sniff out hidden dangers in seemingly innocuous messages.

This shift also raises three major societal concerns:
\begin{itemize}
    \item {\bf AI as Decision-Maker} 

    Entrusting LLMs with security decisions risks diminishing human oversight, opening the door to bias propagation, error amplification, and reduced situational awareness.

    \item {\bf The Centralization Problem}

    Concentrating LLM training within a few entities creates systemic vulnerabilities: biased or manipulated training data can skew user behavior and institutional responses in ethically problematic ways.

    \item {\bf Mono-Technology Dependency }

    Over-reliance on a single AI technology narrows defense diversity and leaves organizations exposed if models fail to generalize to new or adversarial attack vectors.
\end{itemize}

A more robust path embeds LLMs within human-in-the-loop systems, pairing AI-driven scenarios with real-time feedback, verification checkpoints, and interpretive interactions. This hybrid, socio-technical approach blends automated adaptability with critical human judgment, forging a more resilient defense against ever-evolving threats \cite{vishwanath2022weakest}.

\subsection{ Leveraging LLMs for Social Good and Real-Time Threat Moderation}

Attention must also be paid to how LLMs can bolster social safety far beyond the boardroom, intercepting and neutralizing harmful content in real time across public platforms \cite{tao23transparent, tao2024transparent}. By fusing text, image, and audio analysis \cite{vishwamitra2021understandingmeasuringrobustnessmultimodal}, these systems flag everything from disinformation campaigns and hate speech to manipulated media and covert radicalization. On the text side, generative models detect linguistic patterns typical of coordinated influence operations; image forensics uncover deepfakes or doctored visuals; and audio modules recognize hateful or predatory language in voice streams---an especially vital capability for protecting children and other vulnerable populations. In particular, three critical domains stand out:
\begin{itemize}

\item {\bf Hate Speech Detection \cite{hatefulonlinecovid}}

Advanced LLMs can detect coded or veiled language used to bypass traditional keyword filters. By applying reasoning capabilities and prompt engineering, these models can infer the underlying intent, even when malicious messages are obfuscated.

    \item {\bf Election Integrity and Democratic Resilience}
    
   Here, LLM-powered tools monitor social media and online forums for emerging narratives or bot-driven amplification efforts. By analyzing cross-modal cues—such as matching suspicious text with manipulated “evidence” images or audio snippets—these systems can alert moderators and election officials to disinformation surges before they spread. Adaptive simulations then help train civic staff and volunteers, exposing them to progressively sophisticated misinformation tactics so they learn to verify sources, check metadata, and spot coordinated campaigns.

    \item {\bf Child Protection and Online Exploitation Prevention \cite{10903427}}
    
    In this arena, real-time content scanning combines NLP, computer vision, and speech-to-text to detect grooming language, explicit imagery, or illicit solicitations. When a potential threat is identified—say, a user sending age-inappropriate voice messages—the system can automatically flag, quarantine, or route the content to trained human reviewers. Ongoing feedback loops allow the model to learn new slang or evasive tactics used by predators, continually refining its ability to shield minors and support law-enforcement interventions.

\end{itemize}

Embedding LLMs within socio-technical frameworks—anchored by human oversight, continuous feedback loops, and robust ethical guardrails—shifts organizations from reactive moderation to proactive defense, better shielding communities from rapidly evolving digital threats. Such guardrails are essential components of responsible AI system design: models must be transparent and auditable, with built-in safeguards for handling ambiguity. Contextual filters and value-alignment mechanisms, he argued, are indispensable design elements to ensure AI systems uphold social norms and protect user rights.

\subsection{ Role-Based Learning and Operational Integration of LLMs }
Operational strategies play a significant role in integrating LLMs into routine cybersecurity workflows, with a particular focus on industrial control systems and other critical infrastructure sectors. A central concept is role-specific training, in which learning materials and automated assistance are dynamically adapted to individual users’ responsibilities, threat exposures, and data access privileges.

{\bf Role-Specific Contextualization}:
Platforms such as EAGER \cite{cloudflare2016eager}, a crowd-sourced threat-intelligence framework that feeds LLMs with real-time updates and fine-grained context, point to promising approaches for enhancing cybersecurity applications. For instance, a system administrator might receive concise malware-propagation briefings. Alternatively, an operational technology (OT) engineer could be alerted only to SCADA-related vulnerabilities.

By continuously aligning content with actual job functions, LLMs deliver highly relevant guidance that boosts both situational awareness and on-the-job performance.
{\bf Advanced Classification and Retrieval}:
Technical evaluations indicate that decoder-only LLMs \cite{fayyazi2024advancing} excel at classification tasks—such as incident triage and initial threat detection—and can rival more complex architectures. Likewise, Retriever-Augmented Generation (RAG) systems \cite{fayyazi2025proveragprovenancedrivenvulnerabilityanalysis} often outperform traditionally fine-tuned models by dynamically querying external knowledge bases. This plug-and-play updating capability keeps models current without the need for continual retraining—an essential advantage in threat landscapes where new vulnerabilities emerge daily.

{\bf Ethical ``Jailbreak'' Testing}:
While large language models (LLMs) demonstrate significant potential for operationalizing cyber defense strategies, securing the models themselves remains a critical concern. "Jailbreak" serves as a form of ethical penetration testing, which involves crafting adversarial prompts to probe model vulnerabilities, evaluate robustness, and rehearse defenses against potential real-world exploits. This proactive red-teaming methodology strengthens both the resilience of AI systems and the broader organizational security posture.

\subsection{ Discussion}
\begin{enumerate}[1)]
    \item {\it How do we build trust in LLM systems used in high-stakes security decisions?
}

    Trust in large language models (LLMs) must be established through transparency and accountability. Strategies to promote trust include the integration of human feedback loops, enabling users to verify or adjust LLM outputs in real time, and the development of external auditing mechanisms, such as blockchain-based verification systems, to create immutable records of AI decisions. Establishing clear performance benchmarks, including metrics for false positive and false negative rates, is also critical for maintaining stakeholder confidence.

    \item {\it What are the risks of over-relying on LLMs in socio-technical environments?
}

A major concern raised was mono-technology dependency. If organizations rely solely on one LLM architecture or training corpus, they risk systemic vulnerabilities. A pluralistic, layered defense—combining LLMs, human oversight, and diverse tools—was proposed as a more resilient path forward.

    \item {\it Can GenAI effectively detect phishing as attack strategies evolve?
}

Yes, but only with continuous adaptation. An agent-versus-agent approach was identified as a promising strategy, in which generative adversarial models simulate evolving cyberattacks while detection models adapt in parallel. Automated reporting and triage systems were also highlighted as essential components for enabling scalable and effective phishing response mechanisms.

 \end{enumerate}

\section{ LLM Interpretability, Safety, and Security
}
\label{sec:llm-xai}

    



\subsection{Trustworthy AI: Interpretability, Robustness, and Fairness
}

Trustworthy AI research emphasizes three critical aspects: interpretability, robustness, and fairness. Interpretability ensures that AI systems' decisions can be understood by humans, facilitating accountability and trust. Robustness focuses on the reliability of AI models under varying conditions, including adversarial inputs. Fairness addresses the elimination of biases, particularly in high-stakes scenarios such as criminal justice or financial decision-making, where outcomes significantly impact individuals and communities. Together, these principles underpin the development of ethical and dependable AI.

LLMs, characterized by their large number of parameters, encode extensive patterns and associations derived from training data. These parameters, referred to as ``weights,'' determine how the model processes tasks such as answering questions, recognizing correlations, or drawing associations. Each node in an LLM represents a specific encoded concept, though interpreting these in human-understandable terms remains a challenge. Human language, governed by grammar, vocabulary, and context, is inherently complex. While a vocabulary provides a foundation, it does not encompass all meanings, as interpretation depends heavily on usage and contextual cues. This complexity raises the question of whether LLMs truly reflect linguistic meanings or if their representations are fundamentally different from human understanding.

This research endeavor on trustworthy AI examines how concepts are encoded within LLMs. For instance, shallow concepts, such as basic grammatical structures or vocabulary, are often captured in the initial layers of the model, while deeper layers encode more abstract relationships, such as the hierarchy of animal types. This layered structure enables LLMs to process a wide range of tasks, but it also highlights the difficulty of directly mapping their internal representations to human concepts.

Fairness in AI becomes particularly critical in high-stakes scenarios. In criminal justice, for example, models must avoid biases that could disproportionately affect certain demographic groups. Similarly, in financial decision-making, models must provide equitable assessments of creditworthiness to prevent discriminatory practices. The simulations demonstrate how contention arises in such applications, underscoring the need for fine-tuning and robust evaluation to ensure fairness.

While LLMs have made significant advancements in processing and understanding language, challenges remain in aligning their operations with human reasoning and ethical standards. Addressing these challenges is essential for developing AI systems that are both effective and trustworthy \cite{huang2024position}.

\subsection{AI Agent and Cybersecurity
}

The increasing use of AI agents in security-critical applications has introduced novel attack surfaces \cite{fang2024llm,zhan2025adaptive,zhu2025teamsllmagentsexploit}. A striking example of these vulnerabilities is the story of Flexcoin and Poloniex, where NoSQL database flaws were exploited in conjunction with Bitcoin systems, resulting in significant financial losses.

In the case of Flexcoin, attackers exploited a vulnerability in the exchange's implementation of NoSQL databases to manipulate transaction records. This allowed them to siphon Bitcoin funds unnoticed. Similarly, Poloniex suffered from a related exploit where improper query handling led to double-spending issues. These incidents highlight how the intersection of new technologies, such as cryptocurrency and AI, can create cascading security risks. Then, a step-by-step analysis of an AI agent hacking scenario is presented:
\begin{itemize}
    \item Vulnerability Discovery: An attacker identifies a flaw in the AI agent's decision-making or data-handling processes.
    \item Exploiting the Vulnerability: Once the vulnerability is identified, the attacker crafts inputs or actions that exploit it. This could involve injecting malicious code or corrupting the training data of the AI model.
    \item Scaling with Multi-Agent Systems: After a successful breach, attackers often use AI themselves to distribute their efforts across multiple agents. These agents collaborate to avoid detection, target multiple vulnerabilities, and escalate the attack.
\end{itemize}

The landscape of AI security is evolving rapidly, driven by both the increasing use of AI in critical systems and the sophistication of attackers leveraging the same technology. One prominent trend is the integration of AI into attack strategies. Cybercriminals are using AI to automate reconnaissance, generate exploits, and scale attacks more efficiently than ever before. This automation allows attackers to uncover and exploit vulnerabilities at speeds and scales that were previously unattainable. Simultaneously, AI is becoming central to defensive strategies, with organizations employing AI to detect anomalies, predict potential threats, and respond to attacks in real time. However, this has led to an arms race between attackers and defenders, where innovations on one side spur countermeasures on the other. Another significant trend is the emphasis on explainability within AI systems, ensuring that models' decisions are interpretable and auditable. By enhancing transparency, organizations can identify potential vulnerabilities and biases before attackers exploit them. The shift toward integrating adversarial robustness and fairness testing during AI development highlights the importance of building security into the system’s foundation rather than addressing it post-deployment.

Academics and industry leaders are actively addressing the challenges of securing AI systems through innovative approaches. Practical implementations include testing frameworks that simulate real-world attack scenarios to evaluate and improve system resilience. By bridging theoretical insights with practical applications, these efforts aim to ensure that AI systems remain reliable and trustworthy even in the face of evolving threats. The ultimate goal is to create a future where AI not only enhances productivity and decision-making but also operates securely and ethically in diverse, high-stakes environments.

\subsection{AI Security in the Era of LLM: From Model to System
}

From a model and system perspective, AI systems introduce new security threats that go beyond traditional cybersecurity challenges. One significant concern is the ability of language models to inadvertently provide harmful information, such as detailed instructions on constructing dangerous items. While current losses from AI misuse are already concerning, the potential scale of harm remains underestimated due to the variety of ways in which language can be manipulated to elicit unintended outputs. For example, users can subtly rephrase requests to bypass safeguards, exploiting the inherent flexibility and ambiguity of language. Language itself, with properties like synonymy, polysemy, and contextual interpretation, makes it challenging to preempt all harmful uses. For instance, asking ``How can I safely disassemble this device?'' might be benign in one context but malicious in another.

A particularly concerning issue is how language models may be manipulated through ``jailbreaking,'' a method of exploiting their prompts to bypass safety restrictions. Jailbreaking often involves crafting cleverly worded or obfuscated inputs that confuse the model into providing inappropriate responses. As models grow in capability, they also become more challenging to secure as attackers exploit these expanded functionalities. For instance, multi-step queries that initially seem harmless can be chained together to achieve malicious ends, posing a significant challenge for developers to design safeguards that remain effective across diverse use cases.

Moreover, the sheer scale of these models amplifies their potential for misuse. With billions of parameters, models like GPT-4 encode vast amounts of information, making it difficult to identify and isolate problematic responses during training or testing. While language has nuanced properties, AI systems often lack the deeper contextual awareness that humans naturally apply, leading to unintended outputs when faced with ambiguous or adversarial inputs. These challenges highlight the need for a proactive approach to security that combines advanced tools with human expertise.

To address these threats, advanced tools like black-box red-teaming frameworks are being developed to rigorously test AI systems for vulnerabilities. These tools simulate diverse adversarial scenarios, uncovering weaknesses in model responses without requiring direct access to the model's internal structure. One notable development is Autodan-Turbo, an automated tool that achieves high attack success rates (ASRs) across various models by systematically probing for exploitable behaviors. Such tools provide a critical mechanism for identifying and mitigating risks before models are deployed in the wild.

Looking to the future, the focus must shift toward developing robust defense strategies that integrate seamlessly into the AI lifecycle. This includes creating models that not only detect and block malicious prompts but also adapt dynamically to evolving threat patterns. Additionally, interdisciplinary collaboration between AI researchers, linguists, and security experts will be essential for designing systems that leverage a deeper understanding of language properties to anticipate and neutralize misuse. In this way, AI systems can be made safer and more resilient while continuing to deliver their transformative benefits.

\subsection{ Discussion}
\begin{enumerate}[1)]
    \item {\it How Statistics Can Bias Features and Take Effect
}

    Statistics can introduce bias when the data used to train models is unbalanced or incomplete. For example, if a dataset over-represents certain groups, the model may prioritize those features, leading to biased results. This bias can be subtle and challenging to detect, especially when it’s embedded in the feature selection or engineering process, which makes it crucial to carefully assess and balance the data used.

    \item {\it How to Deal with Agents with Different Languages?
}

Agents that work in multiple languages offer advantages like determinism and better control over the system’s behavior. However, integrating different languages presents challenges, such as maintaining linguistic nuances and handling cross-lingual ambiguity. These challenges require careful engineering, but advances in "new natural" systems are making multilingual agents more flexible and effective.

    \item {\it What are the Scientific-Level Challenges for LLMs?
}

A major challenge for LLMs is ensuring sufficient representation of language and knowledge across domains. Another issue is resilience—how models maintain accuracy across varied conditions, such as adversarial attacks or changing environments. We currently lack platforms to measure resilience properly, and developing methods that capture broader knowledge beyond data-centric approaches remains an open problem.

\item {\it What are the Key Challenges and Future Directions for System Agent Security?
}

In system agent security, a major concern is how agents adapt to environmental feedback. In multi-agent systems, collaboration can introduce security risks that must be addressed. Future research should focus on developing secure, collaborative computing systems where agents can safely interact and adapt to dynamic environments.

\item {\it How to Teach the Current AI without a Textbook?
}

Traditional computer science textbooks provide clear structures for teaching, but LLMs don’t fit within this framework. The unpredictable nature of LLMs and their complexity make them difficult to teach using traditional methods. This gap presents exciting challenges for educators and researchers, encouraging the development of new approaches for understanding and applying LLMs.

 \end{enumerate}

\section{Conclusion}
Large language models (LLMs) are increasingly contributing to cybersecurity by extracting actionable insights from unstructured cyber threat intelligence (CTI) reports, with specialized models such as SecureBERT handling domain-specific language. Nevertheless, challenges in reasoning and prompt design persist. In the context of 5G security, LLMs assist in identifying vulnerabilities through automated testing, supporting both protocol-based (top-down) and traffic-based (bottom-up) approaches. These models streamline test case generation, fuzzing, and penetration testing, thereby accelerating vulnerability detection. In security application engineering, LLMs support the development of secure software, with pilot programs training students to apply LLMs to tasks such as code summarization and vulnerability discovery, emphasizing creativity and critical thinking. Although LLMs offer considerable potential, their deployment requires cautious implementation and sustained human oversight.

LLM-powered agents are also transforming business and cybersecurity operations by automating tasks such as incident analysis and data extraction. Techniques, including Retrieval-Augmented Generation (RAG) and agentic LLM frameworks, enhance AI's ability to handle complex queries and support decision-making processes. Cross-domain learning enables LLMs to generalize across cybersecurity tasks; however, challenges such as factual inaccuracies and hallucinations remain. Compared to conventional deep learning models, LLMs offer improved reasoning capabilities but still necessitate human validation. Autonomous agents leveraging Purple Teaming and meta-agent frameworks streamline security operations, with smaller, robust language models increasing reliability and efficiency through reinforcement learning-based data filtering. These developments support the creation of more adaptable, efficient, and trustworthy AI-driven cybersecurity systems.

LLMs are further reshaping cybersecurity practices through adaptive training and real-time threat detection. A shift from traditional awareness programs to resilience-focused training is underway, employing LLMs to simulate personalized phishing attacks based on individual user behavior. Hybrid systems that combine AI capabilities with human feedback are increasingly necessary to mitigate risks such as algorithmic bias and overreliance on automated technologies. In the realm of social safety, LLMs contribute to content moderation by detecting hate speech and promoting child safety in online environments. The ethical design of AI systems—ensuring transparency, accountability, and alignment with societal values—is critical. Operational cybersecurity applications increasingly incorporate role-specific training, real-time threat update frameworks such as EAGER, and enhanced retrieval strategies like RAG. Proactive testing methodologies, including controlled "jailbreak" exercises, are recognized as essential for assessing model robustness and ensuring operational reliability.

Trustworthy AI in cybersecurity emphasizes interpretability, robustness, and fairness to guarantee that AI systems are understandable, reliable, and unbiased, particularly in sensitive domains such as criminal justice and finance. Despite advancements, LLMs with complex architectures continue to face challenges in aligning outputs with human understanding and ethical standards. Emerging vulnerabilities associated with AI agents---exemplified by incidents such as the Flexcoin and Poloniex breaches---highlight the dual role of AI in both cyberattacks and defense mechanisms. Addressing threats such as jailbreak attacks, where subtle input manipulations circumvent model safeguards, remains a priority. Given the scale and flexibility of LLMs, security hardening is increasingly complex; however, methodologies such as black-box red-teaming are proving effective in vulnerability discovery. Future AI systems must integrate dynamic defense strategies and foster collaboration between technical and domain experts to achieve resilient and secure operations.

\section{Acknowledgments}
This report was developed based on discussions generated during a recent workshop on large language models for network security. The authors would like to express their gratitude to the National Science Foundation for its support and to all workshop participants for their valuable contributions, insightful discussions, and engagement throughout the event. Their expertise and perspectives were instrumental in shaping the findings and directions summarized in this report. 

Specifically, \Cref{sec:llm-network} on \textit{LLM Applications in Network Security} was developed based on insightful discussions with Ehab Al-Shaer from Carnegie Mellon University, Hongxin Hu from the State University of New York at Buffalo, Wu-chang Feng from Portland State University, and Junaid Farooq from the University of Michigan-Dearborn.

\Cref{sec:llm-agent} on \textit{LLM Agent and Applications} reflects the valuable insights and contributions of Roberto Rodriguez from Microsoft, Peng Liu from the Pennsylvania State University, Tao Li and Quanyan Zhu from New York University, and Jia Xu from 
Stevens Institute of Technology. 

\Cref{sec:llm-human} on \textit{Socio-Technical Aspects of LLM and Security} was prepared with the benefit of discussions with Arun Vishwanath from Avant Research Group, Hongxin Hu from the State University of New York at Buffalo, Shanchieh Jay Yang from Gonzaga University, and Juntao Chen from Fordham University.

\Cref{sec:llm-xai} on \textit{LLM Interpretability, Safety, and Security} was informed by the discussions with Jia Xu from 
Stevens Institute of Technology, Daniel Kang from the University of Illinois Urbana-Champaign, Chaowei Xiao from the University of Wisconsin, Madison, and Tao Li from New York University.

\addcontentsline{toc}{section}{References}
\bibliographystyle{abbrv}
\bibliography{wksp-llm.bib}

\begin{appendices}

\section{Workshop Organization}
\label{app:workshop}
Dr. Quanyan Zhu, Associate Professor in the Department of Electrical and Computer Engineering at New York University, organized this workshop on October 2-3, 2024. The workshop gathered a distinguished cohort of leading and active researchers from academia, government, and industry. A notable aspect of this event was its keen emphasis on fostering local student involvement, providing a platform for the next generation of cybersecurity enthusiasts to engage with seasoned experts. Tao Li, a PhD candidate, co-organized the workshop and led a student task force responsible for its logistics. The organizing team gratefully acknowledges the support from the NSF Division of Computer and Network Systems and the program director, Dr. Xiaogang (Cliff) Wang.

The workshop was hosted at 370 Jay Street, Brooklyn, NY. More information is available at the workshop homepage: \url{https://nyu-larx.github.io/nsf-llm4security/}. The detailed agenda is as follows. \\

\begin{tabularx}{\linewidth}{@{} l X @{}}
\toprule
\textbf{Oct. 2} & \textbf{Event} \\
\midrule
08:30 - 09:00 & Arrival and Breakfast \\
09:00 - 09:10 & Opening Remarks \\
09:10 - 11:00 & \textbf{Session 1 - LLM Applications in Network Security} \newline
Presenters and Panelists: Ehab Al-Shaer, Hongxin Hu, Wu-chang Feng \newline
Facilitator: Junaid Farooq \newline
Scribe: Tao Li \\
11:00 - 11:30 & Discussions \\
11:30 - 12:30 & Lunch \\
12:30 - 14:20 & \textbf{Session 2: LLM Agent and Applications} \newline
Presenters and Panelists: Roberto Rodriguez, Peng Liu, Tao Li, Jia Xu \newline
Facilitator: Quanyan Zhu \newline
Scribe: Lucile Yang \\
14:20 - 15:00 & Discussions \\
15:00 - 15:30 & Break \\
15:30 - 17:00 & \textbf{Session 3: Socio-Technical Aspects of LLM and Security} \newline
Presenters and Panelists: Arun Vishwanath, Hongxin Hu, Shanchieh Jay Yang \newline
Facilitator: Juntao Chen \newline
Scribe: Yunian Pan \\
17:00 - 17:30 & Discussions \\
\toprule
\textbf{Oct. 3} & \textbf{Event} \\
\midrule
08:30 - 09:00 & Arrival and Breakfast \\
09:00 - 09:30 & Day 2 Opening and Day 1 Summary \\
09:30 - 11:00 & \textbf{Session 4: LLM Interpretability, Safety, and Security} \newline
Presenters and Panelists: Jia Xu, Daniel Kang, Chaowei Xiao  \newline
Facilitator: Tao Li \newline
Scribe: Yuhan Yang \\
11:00 - 11:30 & Discussions \\
11:30 - 12:00 & NSF Remarks \\
12:00        & Lunch Time \\
\bottomrule
\end{tabularx}

\end{appendices}

\end{document}